\documentclass{optica-article}

\journal{opticajournal} % for journals or Optica Open

\articletype{Research Article}

\usepackage{subcaption}
\usepackage{comment}
\usepackage{lineno}
\begin{document}

\title{Resonant enhanced detection of the higher-order modes of a locked cavity}

\author{Ricardo Cabrita,\authormark{1,*} Aaron Goodwin-Jones,\authormark{1} Joris van Heijningen,\authormark{2,3} Pavel Demin,\authormark{1}, Martin~van~Beuzekom,  \authormark{3}
Matteo~Tacca, \authormark{3}
Giacomo Bruno, \authormark{1} 
and Clement Lauzin\authormark{1}}

\address{\authormark{1}Universite Catholique de Louvain, 2, Chemin du Cyclotron, 1348 Louvain-la-Neuve, Belgium\\
\authormark{2}Nikhef, Science Park 105, 1098 XG Amsterdam, Netherlands
\authormark{3}Vrije Universiteit Amsterdam, Amsterdam, Netherlands\\
}
\email{\authormark{*}ricardo.cabrita@uclouvain.be} %% email address is required; see note below about the corresponding author designation

% use {asbstract*} to suppress the copyright line. Copyright information will be added in production

\begin{abstract*}  

Current gravitational-wave (GW) detectors are limited in the amount of circulating power they can reach. Optical absorption in the test masses leads to thermal effects that shift the eigenmodes of the optical cavities, and cause control issues such as parametric instabilities.

Here we experimentally validate a novel technique using optical injection to measure the mode amplitudes within an optical resonator. We use a phase camera, similar to the ones installed at gravitational-wave detectors, in transmission of the cavity, to confirm the mode basis and image modes up to order 10. We showcase as well the capability of the phase camera to determine the optical phase between the carrier fundamental mode and other co-resonating higher-order modes, which can be used for optical suppression of parametric instabilities and automatic mode matching. 

These results highlight the relevance of implementing a similar scheme in current GW detectors to monitor thermal effects.

\end{abstract*}

%%%%%%%%%%%%%%%%%%%%%%%%%%  body  %%%%%%%%%%%%%%%%%%%%%%%%%%

\section{Introduction} 

Optical resonators with high circulating powers 
are crucial in a broad range of fields, such as  phase-contrast electron-microscopy \cite{phaseContrastEM_Mueller2021}, dark matter detection \cite{heinze2024first}, gravitational-wave (GW) detectors \cite{LIGOScientific:2014pky, AdvancedVirgo15}, steady-state microbunching for the production of EUV radiation \cite{liu2024prototype} and even in nuclear fusion with proposals for novel neutral beam heating systems \cite{fiorucci2022overview, photoneutral_Bresteau}. In GW detectors, increased optical powers lead to an increase in detection sensitivity, particularly at high detection frequencies \cite{martynov2019exploring}. For this reason, next-generation GW detectors plan to have circulating powers in the order of the MW in km-long Fabry-perot cavities \cite{design_study_update_et}. Similar optical powers are required in order to perform efficient photoneutralization of deuterium for nuclear fusion \cite{fiorucci2022overview}. In the case of phase contrast electron-microscopy,  hundreds of $\mathrm{GW/cm^2}$ are needed in order to have the laser beam act as a phase plate for the electron beam \cite{phaseContrastEM_Mueller2021}. 

Operation of optical resonators with such high optical powers can suffer from issues with thermal absorption in the optics and power-induced instabilities. One particular case of interest is the use of high optical powers for gravitational wave detection. Gravitational waves are detected using long-baseline Michelson interferometers capable of measuring differential length changes down to $\mathrm{10^{-19}}$~m \cite{LIGOScientific:2014pky}. In such interferometers km-long Fabry-Perot resonators are used to increase the optical path length traveled by photons, thus increasing their sensitivity.

Current detectors were designed to store up to 750 kW of intra-cavity power \cite{LIGOScientific:2014pky}. Operation at this level of circulating power puts strict constraints on the absorption in the optics of the resonator. Even so, absorption in the substrate and coating of the optics, especially at the center where the beam is incident, leads to thermal-refractive effects and changes in the radius of curvature \cite{Rocchi11}. These effects modify the g-factor of the cavity as it warms up. This thermally induced mismatch is currently a limitation in the achievable intracavity power in GW interferometers. Despite existing thermal actuation, methods to sense and accurately control the mismatch remain elusive, with no mode control feedback loop in place \cite{ModeReview}.

High circulating powers in cavities can also lead to other complications, such as three-mode parametric instabilities (PI) \cite{BRAGINSKY2001331}. An optical higher-order mode (HOM) beats with the fundamental mode, exerting radiation pressure force on the suspended mirror, which in turn can drive an acoustic mode of the mirror. If the acoustic mode energy is not dissipated, the ringing of the mode can kick the cavity out of resonance. For this to happen it is necessary that the optical beating between fundamental and higher-order modes is close to the frequency of the acoustic mode in question, that the higher-order and acoustic modes have sufficient spatial overlap and that the system has low loss for the feedback loop that is established.

Existing strategies to control and mitigate PIs include shifting the frequency of the optical HOM away from the frequency of the problematic acoustic mode by thermal tuning \cite{Degallaix:07}, electrostatic damping \cite{ESDampingPI_BlairGras}, or damping the quality factor of problematic acoustic modes \cite{ResDampersPI_Gras}. Another technique involves the suppression of the optical HOM by using an electro-optic modulator (EOM) to inject the same HOM with opposing phase into the cavity \cite{Bossilkov_PIsuppress}. This has been shown for a first-order optical mode. However, this technique is limited due to difficulties in measuring and tracking the complex PI amplitude, especially for high-order modes. In reference \cite{Schiworski2022} the first real-time imaging of a PI, caused by the fundamental and a first-order mode, was made.

An experimental scheme with similar optical injection as the one in \cite{Bossilkov_PIsuppress} and a pinhole photodiode  was simulated and theoretically developed in order to measure second-order modes from mode mismatch in linear and coupled cavities \cite{Goodwin-Jones:23}. In reference \cite{phaseContrastEM_Mueller2021}, and the supplemental material of \cite{schwartz2019phaseContrast},
an EOM was scanned around the mode separation frequency in order to derive a measurement of the waist size in a near concentric cavity for phase contrast electron microscopy. Reference \cite{Beaumont:24} uses an EOM in a similar way, scanning its modulation frequency to measure the HOMs in a v-shaped cavity locked with optical feedback.

It is clear from the  challenges of operating high circulating powers that online monitoring of the cavity eigenmodes is needed. Here, we combine the above mentioned injection scheme with a scanning pinhole phase camera \cite{AgatsumaPC} to measure the amplitude and phase of the eigenmodes of a locked linear cavity. This is done by changing the modulation of an EOM to match the mode frequency of a given HOM. More details in section \ref{opt-inj}. The resulting heterodyne beating between the injected HOM and fundamental mode is then measured in transmission of the cavity using the phase camera, similar to the ones already installed at Virgo \cite{van2020phase}. We demonstrate the capability of this method to monitor the g-factor of a cavity while it is locked. We also analyze the relative power of the different cavity modes taking into account the overlap integral between the fundamental and the sideband higher-order mode to retrieve the calibrated mode powers.

Finally, we  measure the optical phase between the carrier fundamental mode and the injected sideband HOM. The ability to measure the phase makes this technique suitable for use in alignment and mode matching control  \cite{Brown:21, Goodwin-Jones:23, CABRITA_MMPC_sim} and PI suppression via optical feedback \cite{Bossilkov_PIsuppress}. 
 
\section{Experimental set-up}

\begin{figure}[htbp]
\centering\includegraphics[width=0.85\columnwidth]{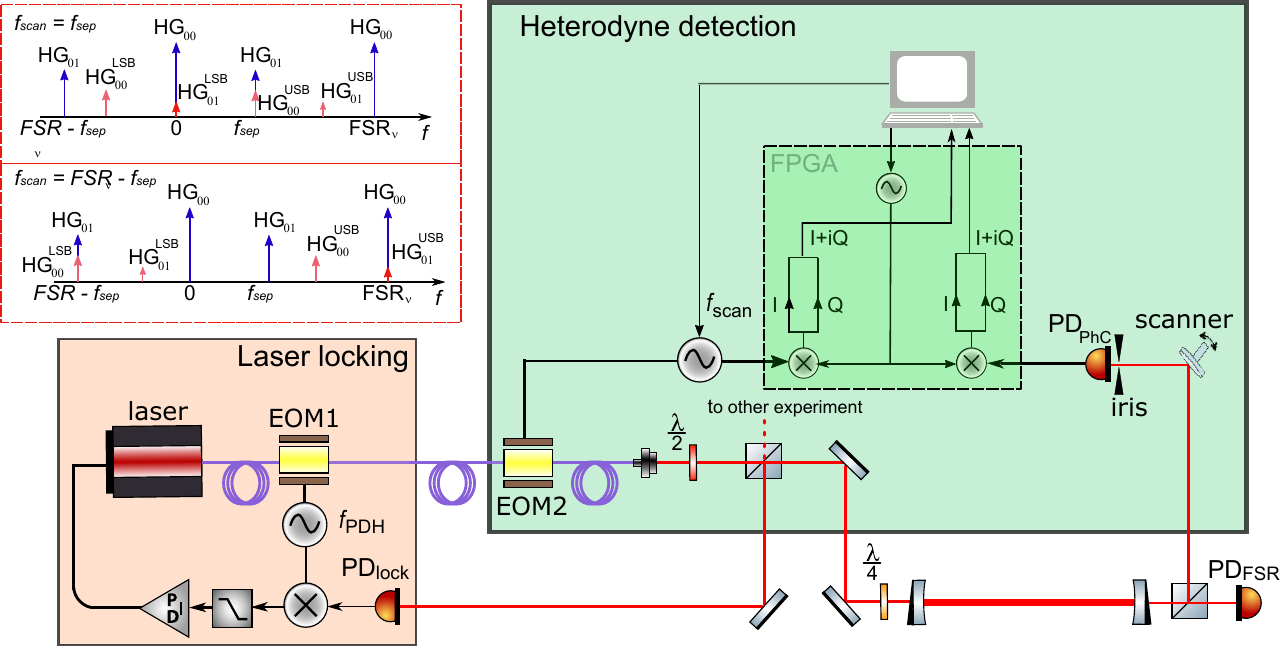}
\caption{Optical schematic of the experiment. The orange colored area shows the optical and electronic components to lock the laser to the cavity resonance \cite{PDHdrever}. The green colored area shows the sideband HOM injection and heterodyne measurement with a phase camera in transmission of the cavity. More details given in sections \ref{opt-inj} and \ref{hetdet-phc}. The top left inset shows a diagram of how the sideband HOMs (red arrows) are injected into the cavity, co-resonating with the carrier (blue arrows). More details given in section \ref{opt-inj}.}
\label{fig:explayout}
\end{figure}

The optical schematic of the experimental set-up is shown in Fig. \ref{fig:explayout}. It is composed of an optical cavity and two heterodyne detection schemes, one in reflection of the cavity and another in transmission. The cavity is composed of an invar bar and two mirrors mounted on K05S2 polaris mounts, separated by  $\approx$320~mm, resulting in a free spectral range (FSR) of $\mathrm{\approx 468.4 MHz}$. The two mirrors have a radius of curvature (RoC) of 500~mm with a nominal reflection of $\mathrm{99.15 \pm 0.15~\%}$, resulting in a nominal finesse $ \mathrm{F \approx 368^{+79}_{-55}}$ and a linewidth $\mathrm{ \Delta \nu \approx 1.27^{+0.23}_{-0.22}~ MHz}$. The laser used is a butterfly packaged DFB laser from Eblana, EP1550-0-DFB-B01-FM, with a linewidth as low as 100~kHz.

The reflection detection scheme, shown in the orange colored area of Fig. \ref{fig:explayout}, is used to frequency lock the fundamental mode of the laser to the cavity using the Pound-Drever-Hall (PDH) locking scheme \cite{PDHdrever}. The error signal is obtained with a first pair of radio-frequency (RF) sidebands at $\mathrm{f_{PDH} = 50~MHz}$, generated by an EOM (EOM1 in Fig. \ref{fig:explayout} ). The carrier and PDH sideband beating is measured at a fast photodiode (150~MHz bandwidth). Throughout the text we call $f_0$ the carrier fundamental frequency which is locked to the cavity.

Note that a spherical optical resonator decomposes the input beam into a series of its own eigenfunctions or mode basis according to complex coupling coefficients $k_{m'n',mn}$, as derived by Bayer-Helms in reference \cite{bayer1984coupling}. Here, we assume the orthonormal Hermite-Gaussian (HG) basis \cite{bayer1984coupling}, in order to allow for some small amount of astigmatism. The modes, in the basis of the cavity, are written here as $\mathrm{HG_{mn}}$, where m and n are the mode numbers of the 1D Hermite-Gauss polynomial in the x and y transverse coordinates, respectively, and $\mathrm{HG_{m'n'}}$ in the basis of the input beam. We assume the input beam is a Gaussian beam perfectly described in its own basis by the fundamental mode $\mathrm{HG_{0'0'}}$. Hence, the transverse input beam decomposed in the basis of the cavity can be written as \cite{bayer1984coupling}:
 \begin{equation}
     U(x,y) = \mathrm{\sum^{\infty}_{m=0} \sum^{\infty}_{n=0} k_{0'0',mn} HG_{mn}(\emph{x},\emph{y})},
 \end{equation}
 where the coupling coefficients are such that:
 \begin{equation}
    \mathrm{\sum^{\infty}_{m=0} \sum^{\infty}_{n=0} \big| k_{0'0',mn} \big|^2 = 1.} 
 \end{equation}

 For simplicity, in the remainder of the text, we omit the input beam indices when writing the coupling coefficients and write them as $k_{mn}$. In this experiment, in order to excite the HOMs, the input beam is significantly mode-mismatched and slightly misaligned with respect to the cavity. 

Finally, note that in linear or Fabry-Perot cavities and in the limit of very small or no astigmatism, HG modes of same $m+n=M$ order resonate together. Thus, when we can't distinguish between same order modes in the cavity we write them as $\mathrm{HG_M}$ with overall coupling coefficient $\mathrm{k_M}$ and refer to them as cavity modes.

\subsection{Optical injection} \label{opt-inj}

 A second pair of sidebands is generated via a second EOM (EOM2 in Fig. \ref{fig:explayout}) driven by a waveform generator at frequency $\mathrm{f_{scan}}$ (Fig. \ref{fig:explayout}). The transverse modes of the generated sidebands will resonate along with the carrier for certain key frequencies. The beating between the carrier $\mathrm{HG_{00}}$ and the co-resonating sideband transverse HOM can be measured in transmission of the cavity, with a heterodyne readout demodulated at $\mathrm{f_{scan}}$. By scanning the EOM modulation frequency, a spectrum of the HOMs of the cavity can be retrieved from the lower sideband (LSB), and another copy of the HOM spectrum at $\mathrm{FSR_{\nu} - HOM_{\nu}}$ can be retrieved from the upper sideband (USB). We exploit this in order to scan the full FSR of the cavity ($\mathrm{\approx 468 ~MHz}$) despite the 250~MHz limit on the acquisition bandwith, by having the LSB probe the first half of the FSR from 0 to 250~MHz, and the USB probe the second half from $\mathrm{FSR_{\nu}}$-250~MHz to $\mathrm{FSR_{\nu}}$.
 
As an example, the upper left inset in Fig. \ref{fig:explayout} illustrates what happens when the EOM is driven at the mode separation frequency of the cavity, $f_{scan} = f_{sep} = FSR_{\nu}/\pi \arccos\left(1-L/RoC\right) \approx 179~\mathrm{MHz}$. In this case, the lower sideband fundamental mode is at $\mathrm{-179 ~MHz}$ from the carrier fundamental mode $\mathrm{HG_{00}}$, while the sideband first order mode is exactly at the carrier fundamental mode frequency, $f_0$. Alternatively, when the EOM is driven at $f_{scan} = \mathrm{FSR_{\nu}-f_{sep}}$, the upper sideband fundamental mode is $\mathrm{\approx 179~MHz}$ away from the next $\mathrm{HG_{00}}$ resonant frequency. In this case, the sideband first order mode frequency is exactly at $\mathrm{f_0 + FSR_{\nu}}$ and thus co-resonanting with the locked carrier. Note that the coupling of the sideband to a $\mathrm{HG_{mn}}$ mode is given by the cavity itself, as is the case for the carrier.
S
The impact of the extra sideband modulation scan on the PDH locking was tested and it was observed that it is largely unaffected except for frequencies exactly at $f_{scan} = n f_{PDH}$, with $n$ a natural number. Note that for longer cavities with an FSR smaller than the RF range, as is the case in GW detectors, this is never an issue.

\subsection{Heterodyne detection with a phase camera} \label{hetdet-phc}

In order to measure the beating between the carrier and the sideband transverse HOM it is necessary to break the orthogonality between the modes. For this purpose, reference \cite{Beaumont:24} uses a masking blade in a translation stage, while in reference \cite{Bossilkov_PIsuppress} a quadrant photodiode,  which works better for first order modes, is used. In reference \cite{Goodwin-Jones:23} the author simulates the use of a clipped photodiode for measuring the beating between the resonant carrier and second order modes.

In this work, we used a custom version of a scanning pinhole phase camera similar to the one described in reference \cite{AgatsumaPC}. In this case a reference beam is not necessary as the beating to be measured is composed of a carrier and a single sideband. The device, composed mainly of a piezo scanning mirror and a fast pinhole photodiode, is shown on the right of Fig. \ref{fig:explayout} in the green area.

The pinhole photodiode clips the beam, breaking the orthogonality between the fundamental and the higher-order mode. The piezo scanner allows moving the beam over the photodiode, either to build an image of the beating or to select the pixel with the best signal-to-noise ratio (SNR). 

The signal is acquired and demodulated by an ALPHA-250 board from Koheron, with four 14~bit ADC inputs of 250~Msps. The photodiode signal is mixed digitally with a local oscillator in the FPGA with a software defined frequency, which is changed as the EOM2 modulation frequency is swept. Digital demodulation allows for acquiring the I and Q quadratures of the optical beating. A computer receives the beating quadratures as a complex number $\mathrm{I +iQ}$ and the amplitude and phase information are simply $\mathrm{ \sqrt{I^2+Q^2} }$ and  $\mathrm{ \arctan{(I/Q)}}$  respectively.
A second ADC input is used with a copy of the EOM RF signal to remove the time dependence on the phase information, as described in reference \cite{AgatsumaPC}.

\section{Results}       

After PDH-locking the laser frequency to the $\mathrm{HG_{00}}$ of the cavity, a DC voltage can be measured at $\mathrm{PD_{FSR}}$, resulting from the resonant carrier $\mathrm{HG_{00}}$ transmitted through the cavity end mirror. Next, $\mathrm{f_{scan}}$ is swept and the AC signal of $\mathrm{PD_{PhC}}$ is digitally demodulated by the FPGA board. For each step of $\mathrm{f_{scan}}$ (from 5 to 249~MHz), the beam is scanned over the photodiode along the x and y cross section and the amplitude and phase are recorded. A full scan with 300 demodulation frequencies and 100 pixels across each x and y cross sections takes about 300 seconds. The amplitude of the measured beating is corrected using the calibrated frequency response of the ADC as well as the calibrated frequency response of the EOM (more details are given in the supplemental document).

\subsection{Heterodyne scan and mode imaging}

\begin{figure}[htbp]
\centering\includegraphics[width=1\columnwidth]{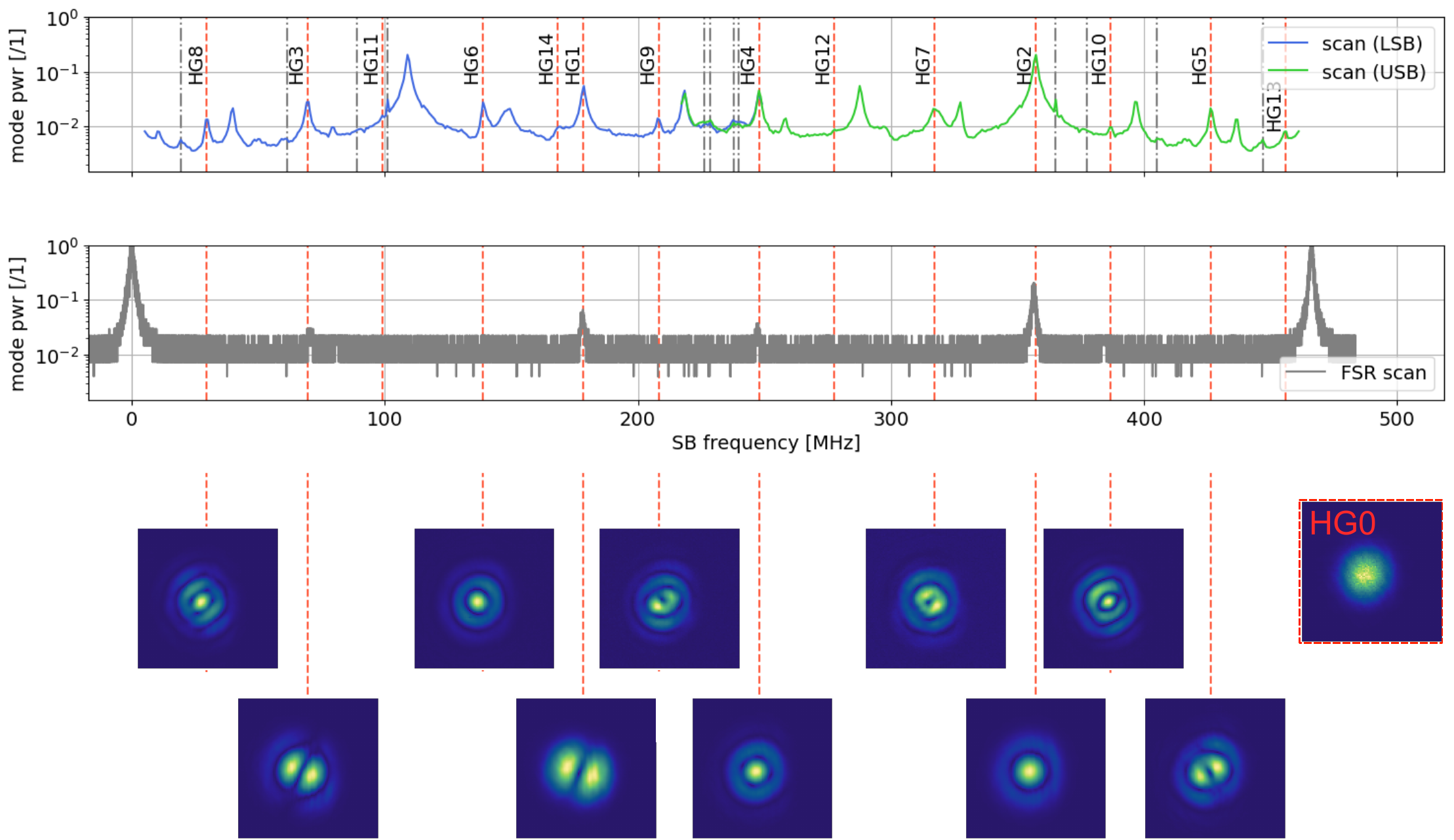}
\caption{Comparison of the resonant heterodyne scan (top panel) with a regular FSR scan (middle panel). The red dashed lines show the injected sideband HOMs. The grey dashed lines tag HOMs from the sideband HOMs of the PDH sidebands.  The frequency scale on the x-axis is obtained directly from the scanning of EOM2. The y-scale calibration is explained in section \ref{rel-int}. Additionally, the amplitude images of all measured modes are shown in the bottom panel. The shape of the fundamental mode is also shown in the red-dashed inset, obtained with 500 kHz sidebands.}
\label{fig:homspectrum}
\end{figure}

Figure \ref{fig:homspectrum} compares the heterodyne amplitude spectrum recorded with the typical cavity FSR scan. The theoretical mode positions, $f_{m+n} = (m+n) f_{sep}$, are shown. For the upper sideband HOMs, the mode positions are given by $FSR_{\nu}- f_{m+n}$. For this reason, the resonant spectrum is duplicated in the figure. The left side for the frequencies probed by the lower sideband (LSB) and the right side for frequencies probed by the upper sideband (USB). The right side is mirrored for scanning frequencies  $FSR_{\nu}- f_{scan}$ in order to match the FSR scan.

Although unintended the PDH sidebands also get modulated by EOM2 and these sideband-of-sideband HOMs will, despite the very small amplitude, also be injected into the cavity. For completion, these HOMs are tagged with dashed grey lines in Fig. \ref{fig:homspectrum}.

In addition to the scan, at each resonant peak mode frequency up to mode order 10, a 100x100 matrix of the beat amplitude is taken. Each image takes about 30 seconds to record. These mode images, shown in the bottom of Fig. \ref{fig:homspectrum}, can be used to verify the mode order and basis. Even if not necessary for our application, an interesting perspective for this work is to combine the images provided by this technique with mode decomposition.

\subsubsection{Extracting the cavity parameters}\label{gfactor_section}

One of the key advantages of this heterodyne scan is that the frequency scale is obtained directly, unlike a cavity scan where an intermediate parameter, such as crystal temperature or piezo voltage is tuned. 

The value of the FSR was found with a fitting procedure using the mode positions obtained with the heterodyne scan. The frequencies of modes 1 through 10 were fitted, using the following equation for modes probed by the LSB:
\begin{equation}
    f_{LSB}(m+n) = \mathrm{FSR_\nu \frac{\emph{m}+\emph{n}}{\pi} arccos(1-L/RoC)},
    \label{eq-lsbpos}
\end{equation}
and, for the modes probed by the USB:
\begin{equation}
    f_{USB}(m+n) = \mathrm{FSR_\nu - FSR_\nu \frac{\emph{m}+\emph{n}}{\pi} arccos(1-L/RoC)},
    \label{eq-usbpos}
\end{equation}
where both L and RoC are the parameters to be fitted, and $\mathrm{FSR_\nu  = c/2L}$ with c the speed of light. Note that the mode positions depend only on the overall cavity g-factor and cannot unambiguously determine the RoC of input and end mirrors. Here we only consider the effective RoC of the cavity mirrors, $\mathrm{(R_{IM}+R_{EM})/2}$. The mode positions were obtained from the measured spectrum by fitting the peaks with a Lorentzian using the Python library lmfit \cite{lmfit}.

A simultaneous fit of both LSB and USB modes was done, having the length and RoC as free parameters. While the RoC was allowed to change within twice the manufacturer tolerances, no bounds were set on the length parameter. The  fitted RoC found was $\mathrm{0.501992~m \pm 70~\mu m}$ and the fitted cavity length $\mathrm{0.321518~m \pm 30 ~\mu m}$. The fitted FSR is then $\mathrm{466.213~MHz \pm 43~kHz}$ and the cavity g-factor is ($\mathrm{g=1-L/RoC}$) $\mathrm{0.35951 \pm 0.00017}$. 

The x-axis of the regular FSR scan in Fig. \ref{fig:homspectrum} was converted to frequency by using the fitted FSR value mentioned above and the distance between consequent fundamental modes.
 
Table \ref{tab:modes} shows the difference between the measured mode frequencies and the fitted mode frequencies, which were obtained with Eq. (\ref{eq-lsbpos}) and Eq. (\ref{eq-usbpos}) using the fitted cavity length and RoC detailed above. The fitted full width at half-maximum (FWHM) of the cavity modes are also shown. The FWHM values fitted for modes of order 7 and 10 were affected by the low SNR of the measured peaks. 

\begin{table}[ht]
    \centering
    \caption{Fitted mode frequencies obtained with Eq. (\ref{eq-lsbpos}) and Eq. (\ref{eq-usbpos}). The measured mode frequencies and FWHM were obtained from the heterodyne scan with a Lorentzian fit of each measured peak. The difference between measured and fitted mode frequencies is also shown.}
    \begin{tabular}{c|c c c c c}
    \hline 
    \hline
       Modes  & Fitted [MHz] & Meas. [MHz] & Diff. [kHz]& FWHM [MHz] & RTL [\%]  \\
    \hline
    \hline
       1 (LSB) & 178.482 & $\mathrm{178.492 \pm 0.015}$ & 10 & $\mathrm{1.75 \pm 0.07}$ & 0.66 \\
       3 (LSB) & 69.429 & $\mathrm{69.394 \pm 0.014}$ & 35 & $\mathrm{1.83 \pm 0.08}$ & 0.77 \\
       6 (LSB) & 138.878 & $\mathrm{138.789 \pm 0.021}$ & 88 & $\mathrm{1.57 \pm 0.09}$ & 0.42 \\
       8 (LSB) & 29.601 & $\mathrm{29.691\pm0.020}$ & 90 &  $\mathrm{1.68 \pm 0.16}$ & 0.57 \\
       9 (LSB) & 208.064 & $\mathrm{208.183 \pm 0.042}$ & 119 & $\mathrm{1.74 \pm 0.26}$ & 0.64 \\
       \hline
       2 (USB) & 109.053 & $\mathrm{109.098 \pm 0.010}$ & 44 & $1.74 \pm 0.044$ & 0.65 \\
       4 (USB) & 218.280 & $\mathrm{218.196\pm0.010}$ & 85 & $\mathrm{1.76 \pm 0.046}$ & 0.67 \\
       5 (USB) & 39.698 & $\mathrm{39.703 \pm 0.016}$ & 5 & $\mathrm{1.70 \pm 0.083}$ & 0.60 \\
       7 (USB) & 148.932 & $\mathrm{148.801\pm0.107}$ & 131 & $\mathrm{4.9 \pm 1.4}$ & NA \\
       10 (USB) & 79.176 & $\mathrm{79.407\pm0.159}$ & 230 & $\mathrm{3.4 \pm 0.68}$ & NA \\
       \hline
       \hline
    \end{tabular}
    \label{tab:modes}
\end{table}

Because of the different spatial distributions different modes can experience different cavity round-trip losses and thus, will experience a different cavity finesse or linewidth. An average finesse of $\mathrm{F = 271 \pm 11}$ is obtained, which is lower than the lower bound on the nominal finesse. The difference could be due to excess optical losses in the cavity due to dust and mirror surface roughness. For this reason, an estimate of the cavity round trip losses (RTL) is also shown in Table \ref{tab:modes}, assuming the mirror reflectance remains at the nominal value of 99.15\%. An average RTL of $\mathrm{0.61 \pm 0.3 \%}$ was computed, where the uncertainty takes into account the manufacturer's uncertainty of 0.15\% in the mirror reflectance.

As pointed out in reference \cite{Beaumont:24}, from the different RTL experienced by different modes, with the help of simulations, a spatial map of the effective cavity RTL could be inferred. Such a map would be with respect to losses at both input and end mirrors as the technique cannot distinguish between the two. 

\subsubsection{Mode power calibration and relative intensities} \label{rel-int}

The y-axis in Fig. \ref{fig:homspectrum} is the ratio of the mode heights with respect to fundamental mode height. For the regular FSR scan this can be done directly. For the heterodyne scan, however, this is not the case.

First, note that the phase camera is not measuring directly the sideband HOM, but the optical beating between the carrier $\mathrm{HG_{0}}$ and the sideband HOM $\mathrm{HG_{M}}$. For a sideband modulated at $\Omega = 2\pi f_{scan}$, resulting in the injection of sideband HOMs of $\mathrm{m+n=M}$ order, the intensity of the optical beating at the phase camera photodiode can be written as (supplemental document):

\begin{multline}
    I_{inj}^\Omega(x,y) = \mathrm{J_0(\beta)J_1(\beta) Re\big[ k_{0} HG_{0} k_{M}^* HG_{M}^* \big] \cos{(\Omega t + \Delta \phi_{M0})} +} \\ \mathrm{J_0(\beta)J_1(\beta) Im\big[ k_{0} HG_{0} k_{M}^* HG_{M}^* \big] \sin{(\Omega t+\Delta \phi_{M0})}},
    \label{eq-beat}
\end{multline}
where $\mathrm{\Delta \phi_{M0}}$ is the phase between the sideband HOM and the carrier fundamental mode, $J_0$ and $J_1$ are bessel functions of the first kind of order 1 and 0, respectively, and $\mathrm{\beta}$ is the modulation depth applied to EOM2. After demodulation, the  $\mathrm{\big( \sqrt{I^2+Q^2} \big)}$ amplitude of the overlap integral is obtained. In order to extract the HOM from the overlap integral, we can divide it by the $\mathrm{HG_{0}}$. Thus, when performing the heterodyne scan, the obtained x and y cross sections at each frequency were divided by the x and y cross section of the $\mathrm{HG_{0}}$ mode. The latter was obtained by modulating EOM2 at 500 kHz, which falls well within the cavity linewidth, thus injecting the $\mathrm{HG_{0}}$ of the sideband into the cavity, which is then measured by the phase camera. 

In order to validate this calibration and since the heterodyne scan cannot probe directly the fundamental mode, we compare the relative intensities of modes order 2 and 1 with their relative intensities from the FSR scan.  For the regular FSR scan, the $\mathrm{HG_2/HG_1}$ ratio is $\mathrm{4.15 \pm 0.19}$, while for the heterodyne scan the $\mathrm{HG_2/HG_1}$ ratio is $\mathrm{3.98 \pm 0.11}$, where the latter is just within the error bar of the first.

Finally, taking into account the  $\mathrm{HG_2}$ and $\mathrm{HG_1}$ FSR scan mode powers, we calibrate the heterodyne scan in terms of mode power as well.

While the heterodyne scan probes HOMs up to order 10, the FSR scan only probes HOMs up to order 4. Note that because of the relatively low finesse of the cavity, there is always some amount of sideband leaking through the cavity, and a direct comparison of the noise floors of the two scans is not straightforward. Additionally, the difference in the oscilloscope and the acquisition board ADCs (8 and 14 bit, respectively) makes it an unfair comparison. 

In the next section, we provide a quantitative analysis of the noise floor of the heterodyne scan and its sensitivity limits.

\subsection{Sensitivity, noise and detection limits}

The amplitude of the injected sideband HOM of order m+n=M will depend on the EOM modulation depth and the (complex) coupling from the $\mathrm{HG_{00}}$ to the HOMs in the cavity basis, $\mathrm{k_{mn}}$. 
Thus, the amplitude of the sideband HOM in transmission of an (approximately) impedance matched cavity and assuming the EOM frequency matches the HOM resonant frequency, can be approximated by (supplemental document):
\begin{equation}
    E^{M}_{SB_{transm}} \approx \mathrm{\frac{2T}{(2T^2+\mathcal{L})} |k_{M}|J_1(\beta)\frac{E_0}{\sqrt{2}}},
    \label{eq:mnsbtransm}
\end{equation}
\\
where $E_0$ is the amplitude of the input beam to the EOM, $T$ is the power transmission of both input and output mirrors and $\mathcal{L}$ the cavity round-trip losses.

We can express the power in the transmitted sideband HOM in decibels relative to the transmitted carrier, dBc:
\begin{equation}
    \mathrm{10 \log_{10}\left(\frac{|{k_{M}} |^2}{|{k_{0}} |^2}\frac{P_{SB}}{P_{carr}}\right) \quad [dBc]}. 
    \label{eq:dBc}
\end{equation}
From a regular FSR scan we can extract the coupling coefficient ratio for a given order M. Additionally,  the sideband to carrier ratio is known from the EOM calibration for a given modulation depth in use. In this way, Eq. (\ref{eq:dBc}) can be used to calibrate the y scale of the noise measurements. This was done using the relation between modes of order 2 measured with the heterodyne scan and the $\mathrm{HG_2/HG_0}$ measured with the FSR scan. The EOM was calibrated for a given modulation depth and over the range of modulated frequencies in order to retrieve the power ratio between sideband and carrier, which already takes into account power lost to higher-order sidebands. The details of this calibration are given in the supplemental document. 

The top panel of Fig. \ref{fig:noisebudget} shows the noise levels calibrated in this way. The thickness of the lines is given by the uncertainty of the calibration, dominated by the uncertainty of the EOM calibration.

Two noise measurements were made. The first was taken by demodulating the photodiode signal at the various frequencies while scanning the resonant carrier $\mathrm{HG_{00}}$ over the photodiode but without the HOM sideband. This measurement allows us to estimate the phase noise carried by the locked carrier, corresponding to the noise floor of the heterodyne scan. The second is the ADC noise, which was measured by demodulating the signal without a photodiode connected to the ADC board. These noise levels set the sensitivity limits that can be achieved. 

\begin{figure}[ht]
    \centering    \includegraphics[width=0.85\linewidth]{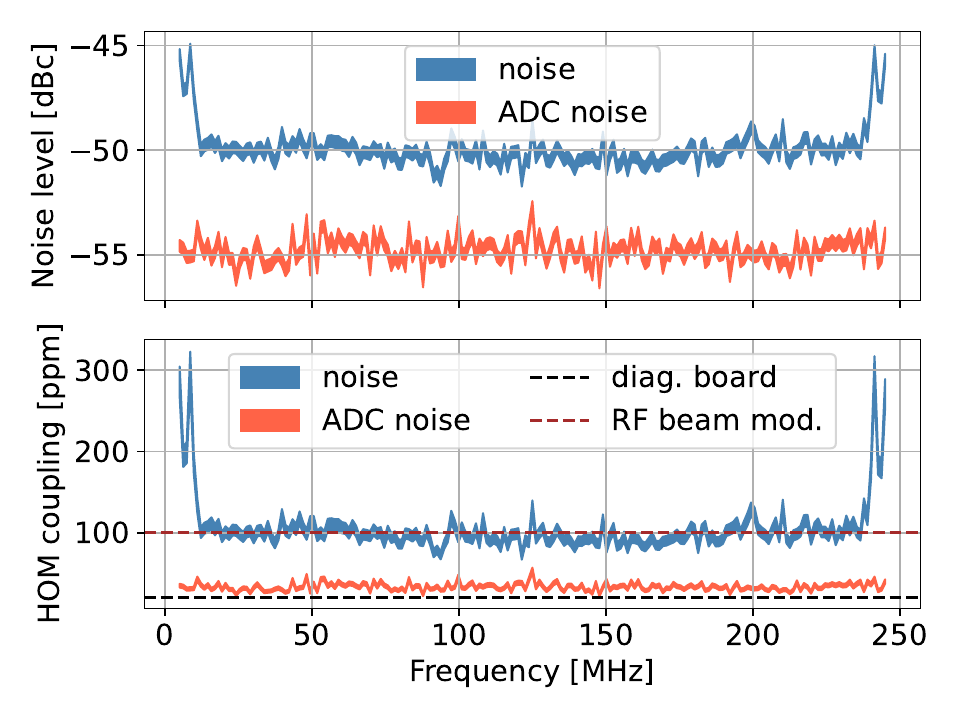}
    \caption{Noise levels of the resonant heterodyne scan in dBc (top panel) and also converted to %and in terms of
    HOM coupling, assuming 10\% sidebands (bottom panel). For comparison, the two best sensitivities reported in \cite{ModeReview} are shown as well: diagnostic board \cite{kwee2007laser} and RF beam modulation \cite{ciobanu2020}.}
    \label{fig:noisebudget}
\end{figure}

The dBc units can be converted to mode coupling $\mathrm{|k_{mn}|}$ for a given value of modulation depth. The bottom panel of Fig. \ref{fig:noisebudget} shows the mode coupling sensitivity limit assuming a measurement made with 10\% sidebands. This is a comparable value to the best sensitivities reported in \cite{ModeReview} for second order modes, such as 20~ppm (RMS) with a diagnostic board based on an optical ring resonator \cite{kwee2007laser} and 100~ppm (RMS) with radio-frequency beam modulation \cite{ciobanu2020}, both shown in Fig. \ref{fig:noisebudget} for comparison.

As the finesse of the cavity increases, the locked laser frequency noise should go down and the blue line in Fig. \ref{fig:noisebudget} should approach the ADC noise level. Note as well that the ADC hardware used here is different from the Virgo phase camera, for which a -61 dBc limit was reported \cite{AgatsumaPC, ModeReview}.

\subsection{Phase sensitive measurements}

In this section, we show both that the phase of the injected sideband HOM can be controlled and that the phase camera can determine the optical phase between the sideband HOM and the carrier $\mathrm{HG_{00}}$.

The phase of the injected HOM can be changed by changing the modulation phase that is applied to the EOM. In this measurement, we sweep the phase of the modulating sine wave applied to EOM2. In turn, the phase of the optical beating between the carrier and the sideband HOM is measured by the phase camera, and given by the $\mathrm{ \arctan{(I/Q)}}$ as explained in section \ref{hetdet-phc}. 

\begin{figure}[ht!]
    \centering
    \includegraphics[width=0.8
    \linewidth]{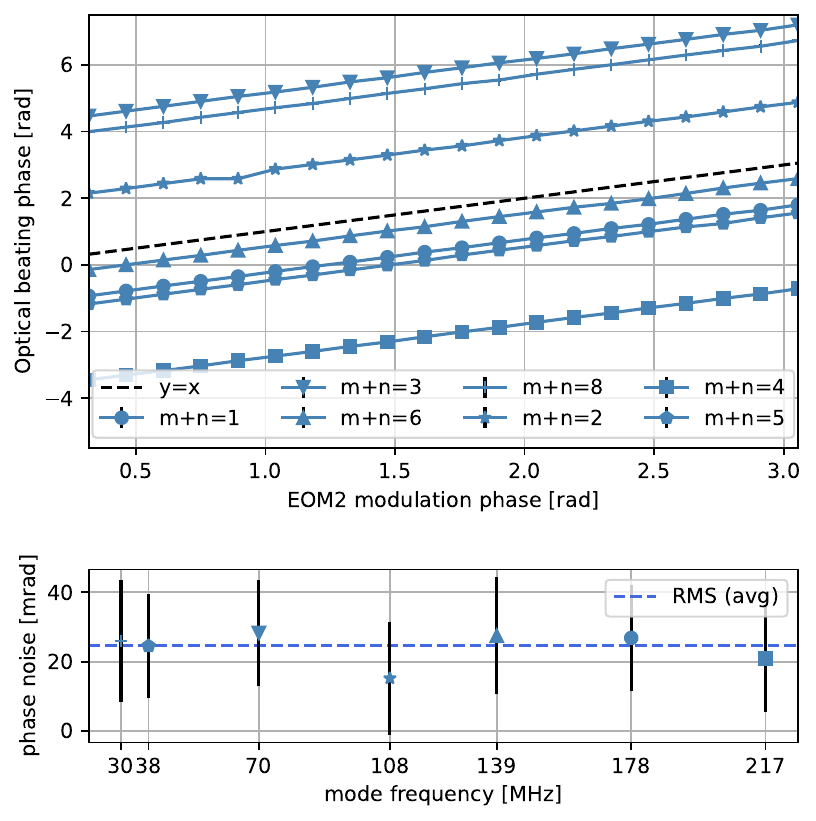}
    \caption{Measured phase of the optical beating between carrier and sideband HOM as function of the modulation phase of EOM2. The bottom panel shows the phase noise as a function of the mode frequency.}
    \label{fig:phasetest}
\end{figure}

The results of these measurements are shown in Fig. \ref{fig:phasetest}, which were repeated for modes up to order 8 (except mode order 7 which had low SNR).

Each datapoint is the average of 100~ms of data and the noise of each datapoint is also stored. The average noise value in mrad for each mode is shown in the bottom panel of Fig. \ref{fig:phasetest}, as well as the overall RMS average which has a value of $\mathrm{\approx 25 ~mrad}$.

The measured beating phase will depend on $\Delta R$, the difference in RoC between the carrier and sideband HOM, the difference in accumulated Gouy phase $\Delta G$, and the optical path difference $\Delta L$, but also in the phase difference $\Delta K$ between the coupling coefficients $k_0$ and $k_M$ and the optical phase difference imprinted on EOM2 $\Delta \phi_{M0}$ (more details on the supplemental document):
\begin{equation}
    \Phi_{beat} = \Delta R - \Delta G + \Delta L + \Delta K + \Delta \phi_{M0}.
\end{equation}

Even though the carrier and sideband are spatially overlapped and same basis modes share the same RoC, in this case we can expect a small but non-zero
$\Delta R$. This is due to the fact that the fundamental and HOM probe different parts of the mirror with slightly different (non-ideal) RoCs. Additionally, due to the difference in frequency, the slight difference in wavevector will also lead to a small but non-zero $\Delta L$. 

As shown in Fig. \ref{fig:phasetest}, the measured optical beating phase follows the modulating phase applied to EOM2, as expected since we only change $\Delta \phi_{M0}$. Note, however, the phase noise is about one order of magnitude higher than that reported in \cite{AgatsumaPC}. This could come from unwanted variations in the other phase affecting parameters. This is especially the case for $\Delta K$. Small mechanical and thermal instabilities in the cavity or in the input beam pointing precision will couple into the alignment and mode mismatch and affect the phase difference between the HOM k couplings. This could also explain the spurious small dip in phase observed for the second order mode in Fig. \ref{fig:phasetest}.

In this way, we have shown the ability to measure the phase of a cavity circulating HOM as well as to control the phase of the injected sideband HOM, both of which are crucial for PI suppression via optical feedback.

\section{Conclusion}

In this work we combined an optical sideband injection scheme with a phase camera heterodyne readout in order to measure and characterize the full eigenmode basis of a locked cavity. Monitoring the changes of the eigenmodes of a cavity as it heats up will be crucial for achieving the MW goal of intracavity power in future GW detectors. This is especially so if this measurement can be used to close the loop in the thermal actuators used in GW detectors.

Additionally, monitoring the linewidth of the injected modes  gives an online map of the RTL as the cavity heats up. It has been observed that increased RTL can be related to mirror damage from circulating powers \cite{liu2024prototype}. Thus, this technique can be used to monitor and map any degradation in the cavity mirrors subject to high optical powers.

We have shown that this set-up can measure HOM couplings $\mathrm{|k_{nm}|}$ as low as 100 ppm. This limit is given by the phase/frequency noise of the locked laser and it should go down with both cavity linewidth and with the improvement of laser frequency noise. Thus, even lower sensitivity limits can be expected if applied in a GW detector.

We have also successfully measured the optical beating phase between the injected sideband HOM and the resonating carrier for modes up to order 8. This shows that a phase camera in transmission of a cavity can be used to perform PI suppression for modes of arbitrary order. The phase sensitive measurements can also be used to automatically correct for misalignments and mode mismatch using the transmission signal.

The authors propose to test this technique at Advanced Virgo ahead of the next observation run O5 which is currently scheduled to start in late 2027.

Finally, this technique can be of interest to areas other than GW detectors, where optical resonators with high intracavity power are crucial. This is the case for phase contrast electron microscopy, photodetachment of anion beams for fusion reactors and steady-state microbunching. 

\bibliography{sample}

%\section{Back matter}

\begin{backmatter}
\bmsection{Funding}
 R.C. is funded under the IISN convention 4.4501.19 “Virgo: physics with gravitation waves” of the Fonds
National de Recherche Scientifique (FNRS). G.B., C.L., A.GJ and R.C. gratefully acknowledge the Region-Waloon (RW) for the financial support under the convention 2410083 “ETOPT” project, Win4Project program. 

\bmsection{Acknowledgment}
The authors thank Herv\'e Laurent and Thomas Schiltz for the technical support. This document was submitted to Virgo and LIGO collaborations with numbers VIR-0454A-25 and P2500129, respectively. The authors thank Ilaria Nardecchia for reviewing the document in both instances.

\bmsection{Disclosures}
The authors declare no conflicts of interest.

%\item When datasets are cited but not submitted as integral supplementary material, they must be cited in the DAS and should appear in the references.

%\bmsection{Data availability} Data underlying the results presented in this paper are available in Ref. [8].

\bmsection{Data availability} Data underlying the results presented in this paper are not publicly available at this time but may be obtained from the authors upon reasonable request.

\bmsection{Supplemental document}
A supplemental document is included. 

\end{backmatter}

\end{document}

\maketitle
\appendix

\section{Phase camera carrier and sideband beating}

The transverse electrical field distribution of a beam can be described by Hermite-Gaussian (HG) modes. These modes form a complete and orthonormal basis to describe any beam under paraxial approximation. Under the assumption of no astigmatism, a mode of a given order m and n can be written as \cite{bayer1984coupling}:
\begin{multline}
    HG_{mn}(x,y,z) = \sqrt{\frac{2}{\pi}}\left(\frac{1}{(2^{m+n}m!n!)} \right)^{1/2} \frac{1}{w(z)}\\
    H_m\left(\frac{\sqrt{2}}{w(z)}x\right)H_n\left(\frac{\sqrt{2}}{w(z)}y\right)e^{-i \big[k\frac{x^2+y^2}{q}-(1+m+n)\Phi_g(q)\big]},
    \label{eq:hgmode}
\end{multline}
where q is the complex beam parameter and $w$ is the size of the beam over z, both defined in reference \cite{Kogelnik1966} in equations 19 and 20 therein. The Gouy phase $\Phi_g$ depends on the order of the mode and is given by $(1+m+n)\arctan{(Re(q)/Im(q))}$.

Since all the results derived here are valid for any value of z, for simplicity we drop the z dependency and write only $HG_{mn}(x,y)$. If we consider the spatial coupling of the input beam electrical field to a cavity, we can write the input laser beam decomposed in the cavity basis as:
 \begin{equation}
     HG_{0'0'}(x,y) = \sum^{\infty}_{m=0} \sum^{\infty}_{n=0} k_{0'0',mn} HG_{mn}(x,y),
 \end{equation}
where we use the Hermite-Gauss (HG) polynomial basis for the spatial profile description of the beam as defined in \cite{bayer1984coupling}. We assume that the input beam is in the fundamental mode of its own basis, which is here denoted by the prime superscript. The complex scattering coefficient from this fundamental mode to a higher-order mode (HOM) in the cavity basis is given by $k_{0'0',mn}$.

If we phase modulate the input beam, at certain key $\Omega$ frequencies, the sideband HOMs with a certain order m+n will co-resonate in the cavity with the fundamental mode of the carrier and the beating. Hence, it is useful to write the phase modulated input beam that will be injected in the cavity. This includes the fundamental mode of the carrier with frequency $\omega$, and for now, a sideband HOM at $\omega+\Omega$ frequency, with $\Omega = 2\pi f_{scan}$:
\begin{equation}
    E_{inj}(x,y,t) =   k_{00}J_0(\beta)HG_{00}(x,y) e^{i(\omega t + \phi_c)}+ k_{mn}J_1(\beta)HG_{mn}(x,y) e^{i((\omega+\Omega)t+\phi_{sb})},
\end{equation}
where $\beta$ is the modulation depth of the sideband, which depends on the voltage applied to the electro-optic modulator (EOM) and $J_0$ and $J_1$ are zero and first order Bessel functions, respectively. Note that for simplicity, the input beam indices were dropped.

Thus, the resonant or injected intensity will be given by:

\begin{multline}
    I_{inj}(x,y) = E_{inj}E_{inj}^* = J_0(\beta)^2|k_{00}|^2|HG_{00}|^2+J_1(\beta)^2|k_{mn}|^2|HG_{mn}|^2 + \\ J_0(\beta) J_1(\beta) k_{00} HG_{00} k_{mn}^* HG_{mn}^* e^{i(\Omega t + \phi_c - \phi_{sb})} + \\
    J_0(\beta) J_1(\beta) k_{mn} HG_{mn} k_{00}^* HG_{00}^* e^{-1(\Omega t + \phi_c - \phi_{sb})},
\end{multline}    
where the $HG_{00} = HG_{00}(x,y)$ and $HG_{mn} = HG_{mn}(x,y)$ abbreviations were made for simplicity.

For the optical beating we are only concerned with the $\Omega$ terms, so we can write the beating intensity as:

\begin{multline}
    I_{inj}^\Omega(x,y) = J_0(\beta) J_1(\beta)  k_{00} HG_{00} k_{mn}^* HG_{mn}^* \cos{(\Omega t+\Delta \phi)} \\
    + i J_0(\beta) J_1(\beta)  k_{00} HG_{00} k_{mn}^* HG_{mn}^* \sin{(\Omega t+\Delta \phi)} \\ + J_0(\beta) J_1(\beta)  k_{mn} HG_{mn} k_{00}^* HG_{00}^* \cos{(\Omega t+\Delta \phi)}\\ - i J_0(\beta) J_1(\beta) k_{mn} HG_{mn} k_{00}^* HG_{00}^* \sin{(\Omega t+\Delta \phi)}, 
\end{multline}
where $\Delta \phi = \phi_c - \phi_{sb}$. Factoring out the cosines and sines and doing some complex number algebra, we finally arrive at:
\begin{multline}
    I_{inj}^\Omega(x,y) = J_0J_1 Re\big[ k_{00} HG_{00} k_{mn}^* HG_{mn}^* \big] \cos{(\Omega t + \Delta \phi)} + \\ J_0J_1 Im\big[ k_{00} HG_{00} k_{mn}^* HG_{mn}^* \big] \sin{(\Omega t+\Delta \phi)}.
\end{multline}

A point of subtlety here is that for linear or Fabry-Perot cavities and in the limit of very small or no astigmatism, modes of the same order m+n are degenerate and resonate together. This will be the case as long as the difference in accumulated Gouy phase in the two axis is less than 2$\pi$ times the cavity linewidth. For this reason, we can talk about cavity modes which will be the sum of all the $HG_{mn}$ with same m+n order. We denote this cavity modes as $HG_M$, using a single index for the sum of the x and y orders and with coupling coefficient $k_M$. In this way, we can write:

\begin{multline}
    I_{inj}^\Omega(x,y) = J_0(\beta)J_1(\beta) Re\big[ k_{0} HG_{0} k_{M}^* HG_{M}^* \big] \cos{(\Omega t + \Delta \phi_{MO})} + \\ J_0(\beta)J_1(\beta) Im\big[ k_{0} HG_{0} k_{M}^* HG_{M}^* \big] \sin{(\Omega t+\Delta \phi_{M0})}.\label{eq-beat}
\end{multline}

Due to the orthogonality of the HG modes, the integrated intensity of equation \ref{eq-beat} over the entire beam is zero. However, the phase camera uses a pinhole photodiode and only measures a small portion of the beam at a time. If the intermodal phase does not vary significantly over the beam, then the integral is non-zero. In appendix B of reference \cite{Goodwin-Jones:23}, the authors workout the integral for the fundamental and second-order modes and plot how the aperture size impacts the measured signal.

\subsection{Demodulation and phase}

A piezo scanner allows for moving the beam over the pinhole photodiode while data is acquired. Each pixel is digitally demodulated with a sine and cosine at $\Omega$ frequency so as to obtain the two quadrature signals:
\begin{align}
    I & = J_0(\beta)J_1(\beta) Re\big[ k_{0} HG_{0} k_{M}^* HG_{M}^* \big] \cos{(\Omega t + \Delta \phi_{M0})}\cos(\Omega t) \\
    Q & = J_0(\beta)J_1(\beta) Im\big[ k_{0} HG_{0} k_{M}^* HG_{M}^* \big] \sin{(\Omega t + \Delta \phi_{M0})}\sin(\Omega t)
\end{align}

Using the rules of cosine and sine multiplication, this results in:

\begin{align}
    I & = \frac{{}J_0(\beta)J_1(\beta)}{2} Re\big[ k_{0} HG_{0} k_{M}^* HG_{M}^* \big] \cos{( \Delta \phi_{M0})} + 2\Omega~term
    \\
    Q & = \frac{J_0(\beta)J_1(\beta)}{2} Im\big[ k_{0} HG_{0} k_{M}^* HG_{M}^* \big] \sin{( \Delta \phi_{M0})} + 2\Omega~term.
\end{align}

The $2\Omega$ terms are removed by a low pass filter and the data is retrieved from the FPGA as a complex number of the form $I+iQ$.

The amplitude signal is given by the modulus of this complex number $\sqrt{I^2+Q^2}$, while the phase is the angle:
\begin{equation}
   \Phi_{beat} = \arctan{\left( \frac{Im\big[ k_{0} HG_{0} k_{M}^* HG_{M}^* \big] \sin{( \Delta \phi_{M0})}}{Re\big[ k_{0} HG_{0} k_{M}^* HG_{M}^* \big] \cos{( \Delta \phi_{M0})}}\right)}.
\end{equation}

Similar to \cite{AgatsumaPC}, from the HG modes we obtain a phase difference from different radius of curvature $\Delta R$, Gouy phase $\Delta G$ and optical path length $\Delta L$:
\begin{equation}
    \Phi_{beat} = \Delta R - \Delta G + \Delta L + \Delta K + \Delta \phi_{M0}, 
\end{equation}
where $\Delta K$ depends on the phase between the coupling coefficients $K_0$ and $k_M$ and $\Delta \phi_{M0}$ is the original phase between the carrier and sideband introduced above. The latter can be changed by changing the phase of the modulating sinewave applied to the EOM.

In this case, since carrier and sideband share the same basis (are spatially overlapped), $\Delta R$ is equal to zero. Different HOMs of the same basis share the same radius of curvature, as implicit in equation \ref{eq:hgmode}. However, when coupled to a spherical cavity there will be a small non-zero $\Delta R$ because the fundamental and HOM probe different parts of the mirror with slightly different (non-ideal) RoC. Due to the difference in frequencies, the slight difference in wavevector will also lead to a small but non-zero $\Delta L$.

\section{Cavity transmitted signal}

Following the derivation in reference \cite{MemoirJerome}, the steady-state circulating power in a cavity on resonance can be approximated by:
\begin{equation}
    P_{circ} = \frac{4T}{(2T+\mathcal{L})^2}P_{in},
\end{equation}
where $P_{in}$ is the input power to the cavity, $\mathcal{L}$ the cavity round trip losses (including mirror losses) and the transmittance of input and end mirrors is $\mathrm{T_{inp} = T_{end} = T}$. Then, the cavity transmitted power can be written as:
\begin{align}
    P_{trans} = P_{circ}T = \frac{4T^2}{(2T+\mathcal{L})^2}P_{in}.
\end{align}
Hence, in terms of amplitude this can be written as:
\begin{equation}
    E_{trans} = \frac{2T}{(2T + \mathcal{L})}\frac{E_0}{\sqrt{2}},
\end{equation}
where $E_0$ is the amplitude of the input beam. The amplitude of the sideband HOMs with order $m+n=M$ which will be injected in the cavity, can be writen as:
\begin{equation}
    E_{SB}^{M} = J_1(\beta)|k_M|E_0.
\end{equation}
The transmitted amplitude of the sideband HOMs can then be written as:
\begin{equation}
    E_{SBtrans}^{M} = \frac{2T}{(2T + \mathcal{L})}J_1(\beta)|k_M|\frac{E_0}{\sqrt{2}}.    
\end{equation}

By scanning the laser frequency around the carrier fundamental mode resonance and measuring the cavity spectrum, the sideband-to-carrier ratio can be measured. This ratio is just a function of $J_1(\beta)/J_0(\beta)$. Additionally, the coupling coefficient ratio $|k_M|/|k_0|$ can also be measured from the FSR scan by taking the ratio in peak heights between an HOM and the  fundamental mode. In this way, we can finally write the sideband in terms of decibels with respect to the carrier as:
\begin{equation}
    \mathrm{10 \log_{10}\left(\frac{|{k_{M}} |^2}{|{k_{0}} |^2}\frac{P_{SB}}{P_{carr}}\right) \quad [dBc]}. 
    \label{eq:dBc}
\end{equation}
We can use the measured signal of the phase camera for a given HOM to calibrate the values returned by the FPGA into dBc units. This was done using the second-order modes and used to display the measured noise in terms of dBc units. The results are shown in section 3.2 of the main document.

\section{Frequency calibration of the phase camera signals}

In order to correctly retrieve the relative heights between different HOMs at different demodulation frequencies, the phase camera signals need to be calibrated in terms of frequency response.

For what concerns the electronics, since the photodiode bandwidth is $\approx$500~MHz and the modulation frequencies used were only up to 250~MHz we only take into account the frequency response of the ADC front-end, which has an acquisition of 250~Msps.

The frequency response of the EOM with respect to the input modulation frequency was also evaluated and found to have a significant impact.

\subsection{ADC frequency response}

\begin{figure}[ht!]
    \centering    \includegraphics[width=0.855\linewidth]{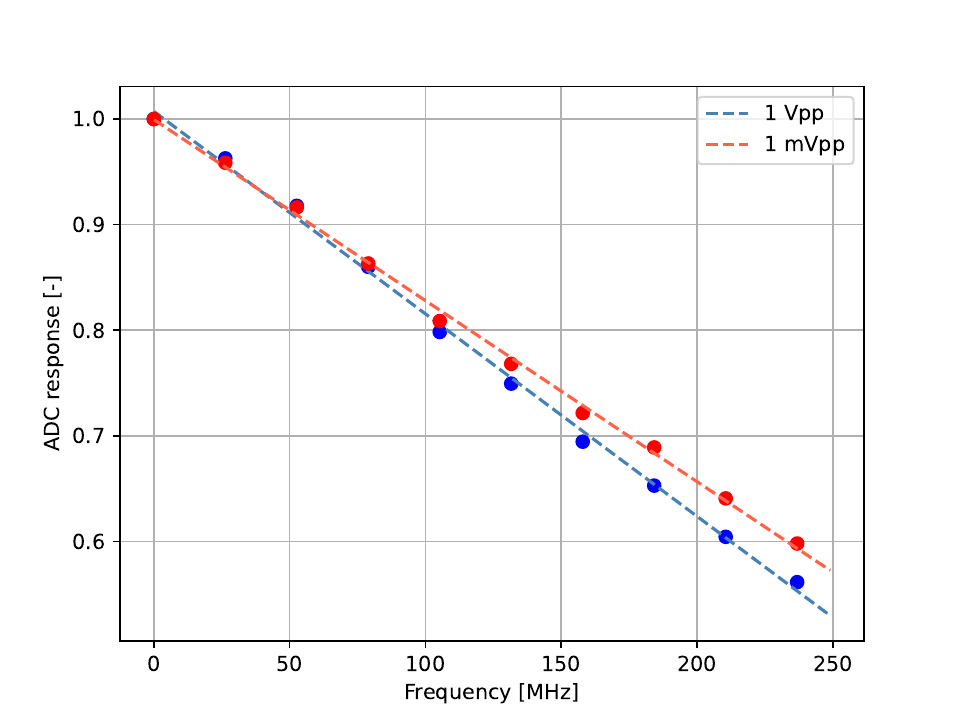}
    \caption{Frequency response of the ADC front-end measured with 1~Vpp and 1~mVpp input signals. Both are normalized with respect to the first data point where no attenuation is expected.}
    \label{fig:adc_freq}
\end{figure}

In order to measure the frequency response of the ADC front-end, the output of a waveform generator was directly connected to the ADC input. The input signal used was a sinewave with frequencies spanning the range of the modulation frequencies used, up to 250~MHz. This measurement was performed with the maximum amplituded of the ADC input, 1~Vpp, and also for small amplitude signals with 1~mVpp. The result of these two measurements are shown in Fig.~\ref{fig:adc_freq}.

The ADC response, defined as $G_{ADC}$ here, was normalized with respect to the first datapoint where no attenuation is expected. It can be seen from Fig.~\ref{fig:adc_freq} that the frequency response for small amplitude signals is slightly different. This was the curve used for the calibration since the optical beating between carrier and sideband HOM is small in amplitude. The correction is applied by dividing the demodulated phase camera spectrum by $1/G_{ADC}$.

\subsection{EOM frequency response}

The EOM frequency response was measured from the ratio of the sideband to carrier peak height. This was measured for different modulation frequencies and with a fixed modulation depth equal to the one used during data taking. The EOM used in the experiment and that we calibrated is a fiber coupled EOM (DC to 300~MHz) from Exail, model MPX-LN-0.1-00-P-P-FA-FA. The driver used to amplify the modulation signals generated by a waveform generator is also from Exail, model DR-VE-0.5-MO (DC to 750~MHz).

\begin{figure}[ht!]
    \centering    \includegraphics[width=0.85\linewidth]{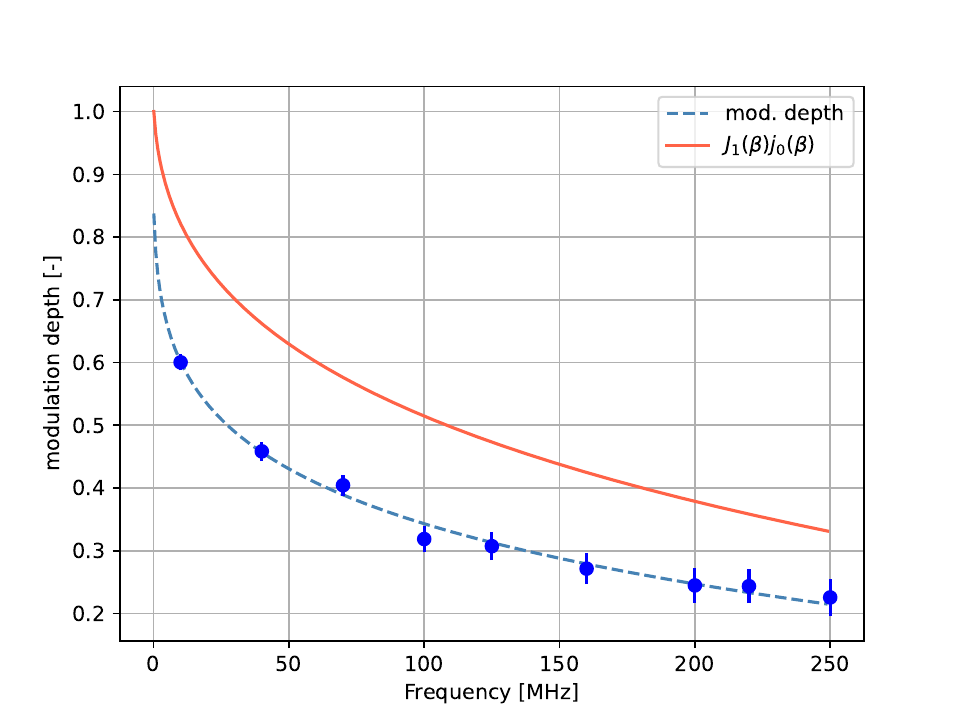}
    \caption{In blue, the measured modulation depth as a function of the modulation frequency. In dashed light blue is the fitted exponential curve. In red we show the actual curve used for correction of the demodulated spectrum. It  is equal to $J_1(\beta)J_0(\beta)$. and was normalized to the value of the first point.}
    \label{fig:eom_freq}
\end{figure}

For each data point measured, the peak-height ratio, $Pk_{SB}/Pk_{carr}$, was evaluated by scanning the laser frequency around the carrier fundamental mode resonance. The corresponding modulation depth was computed as $2*\sqrt{(Pk_{SB}/Pk_{carr})}$ and the obtained values as a function of modulation frequency were fitted with an exponential curve. The measured datapoints and fitted curve are shown in Fig.~\ref{fig:eom_freq}.

In this way, we obtained the frequency dependent modulation depth, $\beta_{eff}(f)$, which is a result of the frequency response of both the EOM crystal and the amplifier driver that amplifies the signals coming out of the waveform generator used for the modulation.

As seen in equation \ref{eq-beat}, the optical beating actually scales with $J_1(\beta)J_0(\beta)$, so the actual correction applied to the demodulated spectrum is $G_{eom} = J_1(\beta_{eff}(f))J_0(\beta_{eff}(f))$. For reference, this correction curve, normalized to the value of the first point, is shown in red in figure \ref{fig:eom_freq}.

\bibliography{sample}